# Single-Spin Asymmetries in Elastic Electron-Hadron Scattering [1]

Andrei Afanasev

*Department of Physics, The George Washington University, Washington, DC 20052*

**Abstract.** This is a short review of the physics of single-spin asymmetries caused by two-photon exchange in elastic scattering of electrons by nucleons and nuclei.

Since the celebrated experiment of Hofstadter in 1950s, electron scattering became a tool in the studies of the structure of a nucleon. In a first Born approximation of the perturbation theory two form factors that describe an electromagnetic current of a nucleon can be related to the cross section of electron-proton scattering according to a Rosenbluth formula [1]. The same form factors enter spin polarization observables in different combinations [2], providing a tool to measure them with scattering of polarized electrons. Measurements of nucleon form factors were done using both methods, and the outcome was a significant discrepancy in electric-to-magnetic form factor ratio; it was summarized, e.g. in Ref. [3].

The elastic cross sections measured in a typical electron scattering experiment are about 10-20% different from the predictions of first Born approximation. The main reason is that an ultra-relativistic electron easily radiates photons, altering the predictions of first Born approximation with logarithmic enhancements. Therefore radiative corrections (RC) need to be applied before one is able to relate experimental data to nucleon form factors. It is the bremsstrahlung diagrams (Fig.1a,d) combined with electron vertex corrections (Fig.1a) that give that largest radiative corrections to cross sections. Procedures to include RC to the data analysis were developed due to Mo and Tsai [4] and applied to the data until recently. RC for the scattering of polarized electrons need to be calculated using an approach beyond soft-photon approximation of Ref.[4]. This program was implemented in Ref.[5], where the authors found that RC to polarization observables are rather small, under a fraction of per cent, for kinematics of Jefferson Lab experiments.

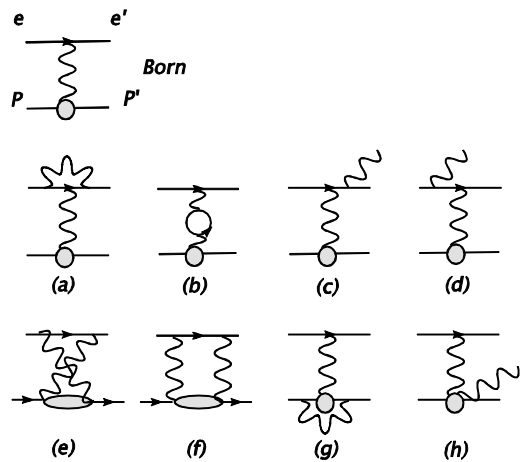

**FIGURE 1.** Feynman diagrams for electron nucleon scattering in the leading order (Born) and next-to-leading-order in electromagnetic coupling. (a) Vertex correction, (b) Vacuum polarization, (c)-(d) Electron bremsstrahlung, (e)-(f) Two-photon exchange, (g) Proton vertex correction, (h) Virtual Compton scattering.

---

Even though the contribution of *two-photon exchange* (Fig.1e,f) is not enhanced by large logarithms, it was first demonstrated by the model calculations in Refs. [6,7] that it has the right sign and approximately the right magnitude to reconcile the outcome of the Rosenbluth and polarization methods. A worldwide program of experimental measurements is currently underway in order to pin down the contribution of two-photon exchange both to the cross section and to the polarization observables of elastic electron-nucleon scattering.

Let us now focus on one distinct feature of the electron scattering beyond plane-wave Born approximation, that is generation of parity-conserving single spin asymmetries (SSA). In plane-wave Born approximation SSA is zero due to time-reversal invariance of the electromagnetic interaction. Corresponding spin-momentum correlation is observed as $\vec{s} \cdot (\vec{k}_1 \times \vec{k}_2)$, where $\vec{k}_{1(2)}$ is an initial (final) momentum and $\vec{s}$ describes the polarization vector of either electron or a proton. It implies that the scattering asymmetry arises due to the polarization component oriented normal to the scattering plane. In the next-to-leading-order in electromagnetic coupling, SSA are generated by the diagrams Fig.1f and, depending on the kinematics, by the virtual Compton process (Fig.1h). The latter only contributes to SSA if the final photon and a nucleon have an invariant mass above the pion production threshold. Therefore the contribution of two-photon exchange can be isolated experimentally by placing a constraint on the invariant mass of the final state.

A similar asymmetry in a pure QED process of electron-muon scattering due to two-photon exchange was calculated by Barut and Fronsdal [8]; later calculations [9] for Moller scattering $e^- + e^- \to e^- + e^-$ also included radiative corrections to the asymmetry. SLAC experiment E158 confirmed the theoretical predictions within the statistical uncertainty [10]: An(exp)=7.04±0.25(stat) ppm vs An(theory)=6.91±0.04 ppm.

Therefore an outstanding issue in theoretical predictions of SSA of elastic electron-hadron scattering is the insufficient information about hadronic structure. On a positive note, experiments on SSA can provide additional information on nucleon structure that is not available from other measurements. Note that in the soft-photon approach [4] to two-photon exchange SSA are zero. Dedicated calculations of SSA in elastic ep-scattering were done by De Rujula et al. [11] for the case of a transversely polarized proton target. As was pointed out in [11], nonzero SSA is due to the absorptive part of the non–forward Compton amplitude for off–shell photons scattering from nucleons.

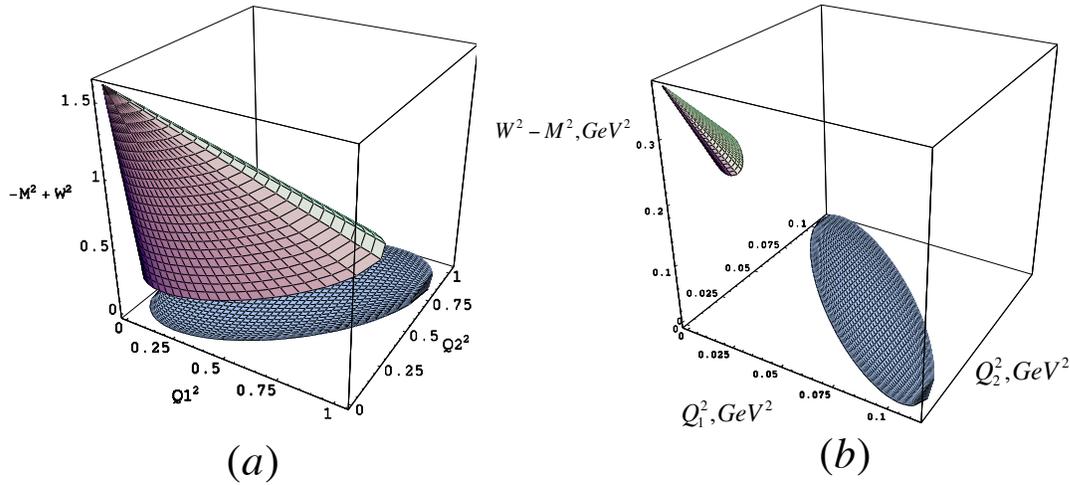

**FIGURE 2.** Kinematic range for the double-virtual Compton scattering amplitude contributing to SSA at fixed electron energy $E_e$ and cms scattering angle $\theta_{cm}$. matching kinematics of beam SSA measurements [12] and [13] (a) $E_e$=855 MeV, $\theta_{cm}$=57 deg; (b) $E_e$=200 MeV, $\theta_{cm}$=145 deg. The gap between two integration regions is due to the pion mass.

Nucleon double-virtual Compton amplitude enters SSA with kinematic factors that are different for beam and target SSA [14]; the result is an integral over the range of kinematic variables, as indicated in Fig.2. Here $W$ is an invariant mass of the intermediate hadronic state, $Q^2_{1(2)}$ are the virtualities the intermediate (space-like) photons, and $M$ is a nucleon mass. The oval-shaped area at $W=M$ in Fig.2 corresponds to the nucleon-only (elastic) intermediate state, and the upper limit of integration in $W$ is constrained by the electron beam energy: $W^2_{max}=M(2E_e+M)$, where we neglected an electron mass for simplicity. If $W=W_{max}$, then $Q^2_{1(2)} \to 0$, therefore it is a quasi-real Compton amplitude that describes this region.

In Ref.[7], the Compton amplitude was modeled by Generalized Parton Distributions obtained from the fits to nucleon form factors. The results are shown in Fig.3 for a target spin asymmetry; corresponding experimental data are anticipated from Jefferson Lab experiment E05-015 [15]. Pasquini and Vanderhaeghen [16] modeled the virtual Compton amplitude by single-pion intermediate states, in the approach that is well justified for the energies below the threshold of two-pion production; they made predictions for both beam and target SSA and obtained good agreement with MAMI data [13].

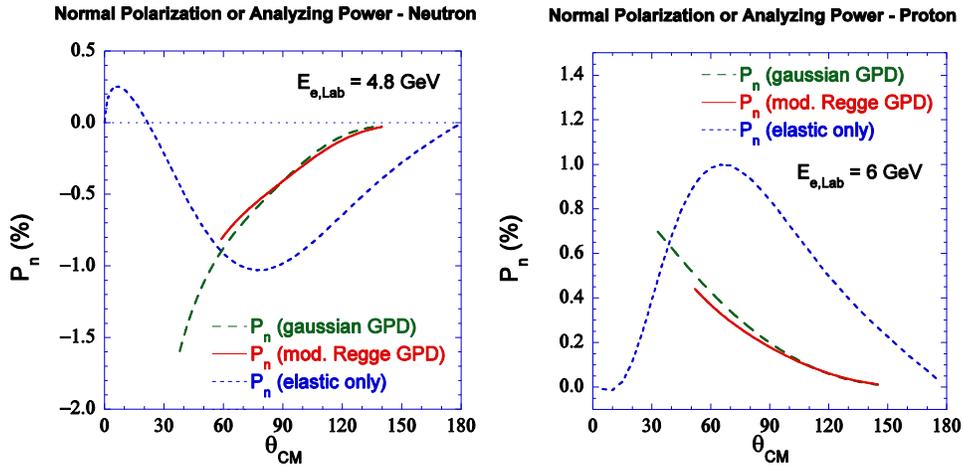

**FIGURE 3.** Transverse target spin asymmetries for elastic electron scattering on a polarized neutron and a proton. Short-dashed curves are contributions from the elastic intermediate state; solid and long-dashed curves are calculations using different models of GPDs, see Ref.[7] for details of the model.

SAMPLE experiment at MIT-Bates [12] was the first to report a measurement of SSA on a nucleon caused by transverse beam polarization; it appeared to be about 3σ away from zero, at a scale of $10^{-5}$. Corresponding SSA were calculated in [14] and showed reasonable agreement (~1.5σ) with the measurement. In [14] the nucleon structure was described by elastic electromagnetic form factors through elastic unitarity. Later work [17] essentially used the same approach to calculations of SSA as in [11,14], but in addition applied a low-momentum expansion, therefore the calculation [17] is limited to low energies of electrons.

Transverse beam SSA for scattering in Coulomb field of a nucleus was calculated a long time ago by Mott [18]. Both in the case of scattering of point-like particles [8,9] and for scattering off the Coulomb center [18] SSA has the following behavior at small scattering angles $\theta_e$:

$$A_n \propto \frac{\alpha \cdot m_e \cdot \theta_e^3}{E_e}, \ \theta_e \ll 1, \tag{1}$$

where $\alpha$ is a fine structure constant and $m_e$ is an electron mass. One can see that for the Coulomb mechanism the asymmetry falls off with beam energy and it is also suppressed for small scattering angles as $\theta_e^3$.

The situation changes dramatically when we consider scattering off a composite target, such as a nucleon or nuclei, that experiences inelastic excitations. We found in a theoretical calculation [19] (confirmed later by a similar derivation [20]) that above the nucleon resonance region the beam SSA (a) does not decrease with beam energy and (b) is enhanced by large logarithms due to exchange of hard collinear virtual photons. The expression for the asymmetry is the simplest in the diffractive regime and small scattering angles, where the virtual Compton amplitude can be related via the optical theorem to total cross section of photo-production on a nucleon by real photons $\sigma_{\gamma p}$ that has little dependence on the energy. As a result, the loop integral for two-photon exchange can be calculated exactly, yielding a simple expression for the asymmetry in the diffractive scattering regime:

$$A_n(diffractive) = \sigma_{\gamma p} \frac{(-m_e)\sqrt{Q^2}}{8\pi^2} \cdot \frac{F_1 - \tau F_2}{F_1^2 + \tau F_2^2} (\log(\frac{Q^2}{m_e^2}) - 2) \cdot Exp(-bQ^2) \qquad (2)$$

with $b$ describing the slope of a non-forward Compton amplitude for the nucleon target, and $F_{1(2)}$ being respectively Dirac and Pauli form factors of a nucleon, and $\tau = Q^2/4M^2$. Comparing Eqs. (1) and (2) one can see that for small angles and high energies the diffractive mechanism may exceed the Coulomb one by several orders of magnitude. We can also apply this approach in the nucleon resonance region, where $\sigma_{\gamma p}$ strongly varies with photon energy. In this case the corresponding factor in the asymmetry is approximately an energy-weighed integral:

$$A_n \propto \frac{1}{\nu_{max}^2} \int_{\nu_{th}}^{\nu_{max}} d\nu \cdot \nu \sigma_{\gamma p}^{tot}(\nu; q_{1,2}^2 \approx 0), \qquad (3)$$

where $\nu$ is a quasi-real photon energy that serves as an integration parameter, and the upper limit $\nu_{max}$ is determined from the electron beam energy.

As long as the electron scattering angle is small - so that the nucleon Compton amplitude can be obtained by extrapolation from the forward limit – this unitarity-based approach gives a good description of experimental data that accompany the measurements on parity-violating electron scattering for a broad range of electron energies, from 1 GeV (Qweak at JLAB ) to 3 GeV (HAPPEX at JLAB) and up to 45 GeV (E158 at SLAC) [10,21]. For the latter, the theoretical prediction based on [19] is $A_n$=-3.2ppm, to be compared with data from SLAC E158 [21] $A_n$=-2.89±0.36(stat)±0.17(syst) ppm, which is about 3 orders of magnitude larger than predictions from Coulomb distortion.

Extension of Eq.(2) to nuclear targets is straightforward, with a ratio of nucleon form factors replaced by an inverse charge form factor of a nucleus [22,23] and $\sigma_{\gamma p}$ replaced by a total photoproduction cross section on a nuclear target. The results appear to be in good agreement with data from HAPPEX and PREX experiments at JLAB [24] obtained on a proton and light nuclei, $^4$He and $^{12}$C. However, a significant disagreement with theory was observed for a high-Z target $^{208}$Pb both in sign and magnitude: $A_n$=0.28±0.25ppm (experiment) against $A_n$≈-8ppm (theory). A possible reason for the disagreement is an effect of Coulomb distortion that grows linearly with a charge of a nucleus and may become significant for this case. A theoretical approach that combines Coulomb distortion and intermediate-state inelastic excitations is required for this case, while experiments with intermediate-mass nuclei, *e.g.* $^{48}$Ca [25] could provide valuable information on transition between different dynamical mechanisms for the asymmetry generation.

In summary, SSA on the nucleon and nuclei provide valuable new information on the nucleon Compton amplitude and multi-photon effects in scattering on nuclei. We look forward to new data coming from the worldwide experimental studies of multi-photon exchange in electron-hadron scattering.

## ACKNOWLEDGMENTS

The author thanks Krishna Kumar and Richard Milner for the invitation to this workshop and stimulating discussions. I am especially thankful to N.P. Merenkov, I. Akushevich, S.J. Brodsky and C.E. Carlson for productive collaboration.